# Realization of a Two-Dimensional Lieb Lattice in a Metal-Inorganic Framework with Flat Bands and Topological Edge States

Wenjun Wu[1][†], Shuo Sun[1][†]*, Chi Sin Tang[2][†], Jing Wu[3], Yu Ma[1], Lingfeng Zhang[1]*, Chuanbing Cai[1], Jianxin Zhong[4], Milorad V. Milošević[5], Andrew T. S. Wee[6,7], Xinmao Yin[1]*

[1]Department of Physics, Shanghai Key Laboratory of High Temperature Superconductors, Shanghai University, Shanghai 200444, China

[2]Singapore Synchrotron Light Source (SSLS), National University of Singapore, Singapore 117603, Singapore

[3]Institute of Materials Research and Engineering (IMRE), Agency for Science, Technology and Research (A*STAR), 2 Fusionopolis Way, Innovis #08-03, Singapore 138634, Singapore

[4]Center for Quantum Science and Technology, Department of Physics, Shanghai University, Shanghai 200444, China

[5]Department of Physics & NANOlab Center of Excellence, University of Antwerp, Groenenborgerlaan 171, B-2020 Antwerp, Belgium

[6]Department of Physics, Faculty of Science, National University of Singapore, Singapore 117542, Singapore

[7]Centre for Advanced 2D Materials and Graphene Research, National University of Singapore, Singapore 117546, Singapore

**\*Emails:** physunshuo@shu.edu.cn; lingfeng_zhang@shu.edu.cn; yinxinmao@shu.edu.cn

**Abstract:** Flat bands and Dirac cones in materials are at the source of the exotic electronic and topological properties. The Lieb lattice is expected to host these electronic structures, arising from quantum destructive interference. Nevertheless, the experimental realization of a two-dimensional Lieb lattice remained challenging to date due to its intrinsic structural instability. After computationally designing a Platinum-Phosphorus (Pt-P) Lieb lattice, we have successfully overcome its structural instability and synthesized it on a gold substrate via molecular beam epitaxy. Low-temperature scanning tunneling microscopy and spectroscopy verified the Lieb lattice's morphology and electronic flat bands. Furthermore, topological Dirac edge states stemming from pronounced spin-orbit coupling induced by heavy Pt atoms have been predicted. These findings convincingly open perspectives for creating metal-inorganic framework-based atomic lattices, offering prospects for strongly correlated phases interplayed with topology.

## 1. Introduction

Electronic flat bands (FBs) in momentum space, when combined with strong electron-electron interactions, are known to foster unconventional superconductivity[1, 2], fractional quantum Hall effect[3, 4], room-temperature ferromagnetism[5, 6], and the formation of Wigner crystals[7, 8]. It has been demonstrated that the strongly correlated states observed in moiré systems such as correlated insulating states and unconventional superconductivity are related to FBs. For example, the magic-angle twisted bilayer graphene displays correlated insulating states at half-filling of FBs, and transitions into a superconducting state when charge carriers are introduced[1, 2]. Nevertheless, up to date, the experimental realization of FBs is still limited to moiré systems and artificial systems. It should be noted that the relatively large unit cell of the moiré systems leads to a low electron density, which may hinder the emergence of the physics typically observed in high-electron-density systems. Hence, it is of significance to construct flat bands in the non-moiré solid materials.

Theoretically, FBs naturally emerge in geometrically frustrated lattices due to quantum destructive interference[9, 10]. One of the notable examples of such frustrated geometries is the Lieb lattice[11], a two-dimensional (2D) edge-centered square lattice, which is also known for its association with the $CuO_2$ planes of cuprate high-temperature superconductors[12]. With each unit cell containing one corner site and two edge-centered sites (Figure 1, a and b), the Lieb lattice exhibits unique electronic properties including FBs, Dirac cones and saddle-point van Hove singularities (VHSs) (Figure 1, a and c), wherein, the nearest-neighbor electron hoppings are taken into account in tight-binding methodology[10]. Additionally, the presence of Dirac cones is associated with high carrier mobility and nontrivial topological properties[13, 14], offering promising prospects for a wide range of applications in electronic devices[15], topological catalysis[16], batteries[17], and supercapacitors[18]. Up to date, the Lieb lattice has primarily been designed in artificial systems due to its intrinsic structural instability, including molecular patterning on metal substrates[19-21], photonic waveguide arrays[22, 23], cold atom systems[24], covalent-organic frameworks (COFs)[25, 26] and metal-organic frameworks (MOFs)[27]. Nevertheless, sustained difficulties in the experimental realization of the Lieb lattice in real solid materials have limited an in-depth exploration of its exotic quantum states that may carry strong implications for both fundamental scientific research and practical applications. Notably, the physical and chemical properties of Lieb-like lattices constructed by COFs or MOFs are easily tunable via strain engineering techniques, doping, and by changing different organic ligands, but the relatively large hopping distances and the complexity of multiple hopping parameters considered in COFs and MOFs have greatly hampered the study of their electronic properties[25-27]. Meanwhile, the physical and chemical properties of Lieb lattices based on metal-inorganic frameworks (MIFs) remained widely elusive. Recently, the successful realization of buckled Lieb lattices (Sn-Al) using molecular beam epitaxy (MBE), which relies on the interaction between specific atoms and a metal surface, has sparked significant interest[28]. However, the characteristic electronic properties of the Lieb lattice including FBs, Dirac cones and VHSs, have not been found in the Sn-Al lattice. Continued efforts are thus needed to synthesize the high chemical stability of novel Lieb lattices and uncover their advanced electronic and topological properties.

Under specific conditions, a Lieb lattice can be epitaxially formed, as illustrated in Figure 1a, facilitated by strong interactions between atoms A and B, driven by their distinct electronegativity[29]. The A atoms are chosen for their high chemical stability to occupy the hollow sites of the B atoms. Meanwhile, the C atoms act as a supportive substrate and provide the foundation for epitaxial growth. In this case, platinum is selected as the A atoms due to its exceptional stability, while phosphorus is chosen as the B atoms for its strong affinity with metals, forming coordination bonds. Gold serves as the substrate layer for its stability and conductivity, thereby bearing in mind the prospect of using Lieb lattices in electronic devices.

With above choices made, we report here the successful *ab initio* computational design of the Platinum-Phosphorus (Pt-P) Lieb lattice based on a MIF, which possesses quintessential properties including FBs, Dirac cones, and VHSs in its electronic band structure. Such Pt-P Lieb lattice was then prepared on a Au(111) substrate via MBE. Subsequent thorough experimental characterization including atomically-resolved scanning tunneling microscopy (STM) imaging and high-resolution scanning tunneling spectroscopy (STS) fully corroborates the structural morphology and the presence of electronic FBs of the Pt-P Lieb lattice, as predicted by theoretical calculations. We further argue the existence of a topological Dirac edge state in this Pt-P Lieb lattice due to the strong spin-orbit coupling (SOC) effect imparted by the heavy Pt atoms. These results unequivocally prove the potential that MIF-based Lieb lattices hold for delving into FB physics and the exploration of nontrivial topological phenomena, up to the standards of technological needs.

## 2. Results

### 2.1. Crystal structures and electronic properties of the Lieb lattice

Figure 2, a and b, displays the atomic model of Pt and P adlayers on a Au(111) substrate. P (cyan) and Pt (orange) atoms form the 1 × 1 and √2 × √2 square-like lattice, respectively, such that the Pt atoms are located at the four-hollow sites of P atoms. As shown in Figure 2c, with increasing interlayer distance, the interlayer binding energy ($E_b$) exhibits a parabolic-like decrease, followed by a linear increase. Our theoretical simulations reveal that the energetically ideal interlayer distances are $d_1$ = 0.8 Å and $d_2$ = 2.2 Å, for which the Pt (P) atoms are arranged in a square-like lattice with the nearest Pt-Pt (P-P) distance of 3.8 Å (2.2 Å). Therefore, Pt and P adlayers form a 2D Lieb lattice, different from the Sn-Al lattice relying on the Sn adlayer and the topmost Al atoms as realized in Ref. 28.

To investigate electronic properties of the 2D Pt-P Lieb lattice, we compare the band structures without and with SOC effects included, as depicted in Figure 2, d and e, respectively. Although the strong SOC effect from heavy Pt atoms leads to prominent changes to several energy bands, distinctive band features in the form of FBs, VHSs and Dirac points that characterize the Pt-P Lieb lattice remain present. Specifically, seven nearly dispersionless FBs occupying most of the Brillouin zone are distributed above and below the Fermi level ($E_F$). These FBs (dashed purple squares in Figure 2, d and e) give rise to distinctive peaks in the computed projected density of states (see Figure S1), and consist primarily of contributions from the P 3$p$ orbitals and Pt 5$d$ orbitals. Meanwhile, the VHS (dashed red circle) is manifested in much flatter dispersion along X-M as compared to the direction X-Γ (see Figure S2). This suggests that the quadratic contribution along X-M is substantially reduced. Intriguingly, the VHSs located near $E_F$ can serve as a potential carrier reservoir, thereby enhancing the superconductivity and catalytic properties of the host material[30, 31]. Additionally, strongly dispersive Dirac cones (dashed blue and green circles) located at the Y-point near $E_F$, may present nontrivial topological properties when the SOC effect is present as further discussed thereafter.

## 2.2. Experimental realization of Lieb lattices in a MIF

With the encouraging results from theoretical calculations, epitaxial growth of P and Pt adlayers on a Au(111) substrate has been systematically carried out. After the deposition of 1 monolayer of P onto a clean Au(111) surface and subsequent annealing at 300 °C (see Figure S3), Pt atoms were deposited, and annealed at 300 °C for 80 minutes (Figure 3a). From the magnified high-resolution STM image in Figure 3b, one can clearly see the bright region formed. The measured distance between P (Pt) adlayers and Au(111) surface is ~2.2 Å (~3.0 Å), corresponding to the theoretical calculations (see Figure 2c, Figures S3 and S4). Figure 3c represents an atomically resolved STM image of one such region, revealing that the short-range ordered Lieb lattice with convincingly square-shaped geometry is formed there. Some point defects are visible as well (dark pits in Figure 3c), which in fact may optimize the charge carrier concentration and scattering of high-frequency phonons, and thus improve some materials' properties[32]. The differential conductance curve (d$I$/d$V$) measured by STS for the sites within the Lieb lattice of Figure 3c is displayed in Figure 3d. Two sharp peaks proportional to density of states (DOS) are observed, one at $V \approx +2.66$ V and one at $V \approx +1.83$ V, corresponding to FB1-2 (~$E_F$ +2.98 eV) and FB3 (~$E_F$ +2.12 eV) in the calculated band structure (see Figure S1), respectively. However, experimental measurements do not reveal the expected FB peaks near and below $E_F$. This discrepancy is likely due to the influence of the Au substrate. Incorporating the effects of the Au substrate into our calculations indicates that it predominantly affects the electronic bands near and below $E_F$ (see Figure S5). This substantially alters the band structures in this energy range, resulting in the elimination of the FB features near and below $E_F$. Consequently, only two distinct FBs located well above $E_F$ remain observable.

Further STM simulation was conducted based on an optimized geometric model with the simulated STM image based on the Tersoff-Hamann method within constant current mode shown in Figure 3g[33]. In this case, with the bright dots representing the Pt atoms, it is in good agreement with the experimental results (Figure 3, c and e) and crystal structures (Figure 3f). The distance between adjacent Pt atoms is measured to be ~4.0 Å (see Figure S4), close to the Pt-Pt bond length of ~3.8 Å within structural optimization. This further supports the reliability of our theoretical model.

## 2.3. Nontrivial topological properties of the Pt-P Lieb lattice

To further explore the nontrivial topological properties of the Pt-P Lieb lattice, we have conducted calculations to investigate its surface and edge states, in particular with the important role that SOC of the heavy elements plays in the band structures near $E_F$. Nontrivial topological effects such as band opening and band inversion are expected to take place[34], further enabling exotic quantum phenomena[35]. This is particularly true in the present case, where the Pt atom as a heavy elemental component has associations with triply degenerate semimetals and multifold degenerate semimetals[36, 37]. In our calculations, the two Dirac cones at the Y-point near $E_F$, as seen in Figure 2d, are particularly interesting. Hence, with the SOC effect considered, there are bandgap openings at the Dirac cone located at ~$E_F$ -0.18 eV and ~$E_F$ +0.26 eV by about ~38 and ~6 meV, respectively (see Figure S6). These are direct signatures of nontrivial topological properties. Figure 4, a and b, provides further details of the energy dispersion function along the (100) surface that corresponds to the Dirac cone below and above $E_F$, respectively. In each case, the edge states are clearly visible as indicated by the red contours in Figure 4, c and d. Interestingly, Figure 4c exhibits a Dirac-cone edge state (in red) at $E_F$, thereby facilitating the flow of electrons from the valence band to the conduction band. Figure 4d shows two edge states (in red) that connect both the bottom and top bands. Even though the edge states are located above $E_F$, their locations can be tuned to the $E_F$ position via strain engineering[38], doping[39], multilayer stacking[40], and applying electric fields[41]. This further enhances its potential for applications in devices, catalysis and energy storage. To verify topologically nontrivial edge states of the Lieb lattice, we calculated the $Z_2$ topological invariants using the WannierTools package[42]. The Wilson loop method is employed to track the evolution of the Wannier charge centers[43], which confirmed the nontrivial topological properties of the Lieb lattice with a nonzero $Z_2 = 1$ (see Figure S7).

## 3. Conclusion

In summary, we have successfully designed a 2D Pt-P Lieb lattice based on first-principles calculations, where unique electronic properties such as FBs, Dirac cones, and VHSs have been anticipated. Based on that in *silico* design, a 2D Pt-P Lieb lattice has been synthesized using MBE, its structure has been validated by STM, and electronic FBs have been detected by STS. Based on the theoretical calculations, we have further predicted the existence of a nontrivial topological Dirac edge state in this Pt-P Lieb lattice. These findings provide a broad platform for design, fabrication, and a comprehensive understanding of the physical and chemical properties of Lieb lattices and will stimulate further theoretical and experimental research on MIF-based Lieb lattices. In other words, our successful design and fabrication of a MIF-based Lieb lattice opens new frontiers in the exploration of special electronic, magnetic, optical and topological properties of high-quality and stable Lieb lattice structures, providing opportunities for their applications in advanced (opto)electronic devices, catalysis, and energy-related technologies.

## 4. Methods

### 4.1. Tight-binding calculations

We constructed a tight-binding model with $D_{4h}$ symmetry, where two edge-centered (A/C) and one corner (B) sites have a coordination number of two and four, respectively. The electronic properties of the Lieb lattice can be calculated from the following tight-binding Hamiltonian:

$$H = \sum_i \epsilon_i c_i^\dagger c_i - t \sum_{\langle i,j \rangle} \left( c_i^\dagger c_j + H.c. \right) - t' \sum_{\langle\langle i,j \rangle\rangle} \left( c_i^\dagger c_j + H.c. \right)$$

where $\epsilon_i$ is the on-site energy of site $i$; $t$ and $t'$ indicate nearest-neighbor $\langle i,j \rangle$ and next-nearest-neighbor $\langle\langle i,j \rangle\rangle$ hopping constants, respectively; $c_i^\dagger$ and $c_i$ are the creation and annihilation operators of an electron on site $i$, respectively. The on-site energy difference between B and C sites is $\Delta E = \epsilon_B - \epsilon_C$. The Lieb lattice is studied with $t'$ and $\Delta E$ equal to zero, and $t$ is set with a negative value.

### 4.2. First-principles calculations

First-principles calculations based on density functional theory (DFT) were undertaken by using the Vienna ab initio simulation package (VASP)[44], adopting the projector-augmented wave method[45]. The exchange-correlation functional based on the generalized gradient approximation (GGA) in the form of the Perdew-Burke-Ernzerhof (PBE) functional[46], was used for both structural optimizations and electronic structure calculations. The energy cutoff was set at 500 eV.

The density of *k*-point grids in the Brillouin zone is 11 ×11 ×1. The convergence criteria for the structural relaxation were that the force and total energy were set to 0.001 eV/Å and $10^{-6}$ eV, respectively. The Wannier90 package was used to construct Wannier tight-binding models[47], and the WannierTools package was used to calculate the surface spectral functions by using the iterative Green's function method[42]. The scanning tunneling microscopy simulation was performed by the Tersoff-Hamann method in a constant current mode[33].

### 4.3. Sample preparation and characterization

The experiments were carried out using a UHV−STM conjoined molecular beam epitaxy system. A clean Au(111) surface was prepared by repeated Ar$^+$ bombardment (1.5 kV, 5 ×10$^{-5}$ mbar) and subsequently annealed at 270 ℃. P atoms were deposited on a Au(111) surface and subsequently annealed at 300 ℃. Pt atoms were then deposited (10.0 Å 20 min + 10.2 Å 30 min) and annealed at 300 ℃ for 80 min. All the scanning tunneling microscopy and spectroscopy (STM/STS) measurements were carried out at 77 K with the bias voltage applied to the sample.

## Data availability

All data supporting the findings of this study are available within the paper and its Supporting Information, or available from the corresponding authors upon reasonable request.

## Acknowledgements

Wenjun Wu, Shuo Sun and Chi Sin Tang contributed equally to this work. This work was supported by National Natural Science Foundation of China(Grant Nos.52172271, 12374378, 52307026, 12374046),the National Key R&D Program of China (Grant No. 2022YFE03150200),Shanghai Science and Technology Innovation Program(Grant No.

22511100200, 23511101600). J.W. acknowledges the Advanced Manufacturing and Engineering Young Individual Research Grant (AME YIRG Grant No. A2084c0170) and the SERC Central Research Fund (CRF). C.S.T. acknowledges the support from the NUS Emerging Scientist Fellowship. S.S. acknowledges support from Natural Science Foundation of China (Grant No. 12304199), Science and Technology Commission of Shanghai Municipality, the Shanghai Venus Sailing Program (Grant No. 23YF1412600).

## Conflict of Interest

The authors declare no competing interests.

## References

[1] Cao Y, Fatemi V, Fang S *et al.* Unconventional superconductivity in magic-angle graphene superlattices. *Nature*. 2018; **556**(7699): 43-50.
[2] Cao Y, Fatemi V, Demir A *et al.* Correlated insulator behaviour at half-filling in magic-angle graphene superlattices. *Nature*. 2018; **556**(7699): 80-84.
[3] Zibrov A, Spanton E, Zhou H *et al.* Even-denominator fractional quantum Hall states at an isospin transition in monolayer graphene. *Nature Physics*. 2018; **14**(9): 930-935.
[4] Léonard J, Kim S, Kwan J *et al.* Realization of a fractional quantum Hall state with ultracold atoms. *Nature*. 2023: 1-5.
[5] Zhao C, Xu Z, Wang H *et al.* Carbon-doped boron nitride nanosheets with ferromagnetism above room temperature. *Advanced Functional Materials*. 2014; **24**(38): 5985-5992.
[6] Bouzerar G. Flat band induced room-temperature ferromagnetism in two-dimensional systems. *Physical Review B*. 2023; **107**(18): 184441.
[7] Li H, Li S, Regan EC *et al.* Imaging two-dimensional generalized Wigner crystals. *Nature*. 2021; **597**(7878): 650-654.
[8] Shayegan M. Wigner crystals in flat band 2D electron systems. *Nature Reviews Physics*. 2022; **4**(4): 212-213.
[9] Regnault N, Xu Y, Li M-R *et al.* Catalogue of flat-band stoichiometric materials. *Nature*. 2022; **603**(7903): 824-828.
[10] Springer MA, Liu T-J, Kuc A *et al.* Topological two-dimensional polymers. *Chemical Society Reviews*. 2020; **49**(7): 2007-2019.
[11] Lieb EH. Two theorems on the Hubbard model. *Physical review letters*. 1989; **62**(10): 1201.
[12] Schilling A, Cantoni M, Guo J *et al.* Superconductivity above 130 k in the hg–ba–ca–cu–o system. *Nature*. 1993; **363**(6424): 56-58.
[13] Hasan MZ, Kane CL. Colloquium: topological insulators. *Reviews of modern physics*. 2010; **82**(4): 3045.
[14] Lv B, Qian T, Ding H. Experimental perspective on three-dimensional topological semimetals. *Reviews of Modern Physics*. 2021; **93**(2): 025002.

[15] Breunig O, Ando Y. Opportunities in topological insulator devices. *Nature Reviews Physics*. 2022; **4**(3): 184-193.

[16] Li G, Fu C, Shi W *et al.* Dirac nodal arc semimetal PtSn$_4$: an ideal platform for understanding surface properties and catalysis for hydrogen evolution. *Angewandte Chemie*. 2019; **131**(37): 13241-13246.

[17] Liu J, Wang S, Sun Q. All-carbon-based porous topological semimetal for Li-ion battery anode material. *Proceedings of the National Academy of Sciences*. 2017; **114**(4): 651-656.

[18] Luo H, Yu P, Li G *et al.* Topological quantum materials for energy conversion and storage. *Nature Reviews Physics*. 2022; **4**(9): 611-624.

[19] Slot MR, Gardenier TS, Jacobse PH *et al.* Experimental realization and characterization of an electronic Lieb lattice. *Nature physics*. 2017; **13**(7): 672-676.

[20] Drost R, Ojanen T, Harju A *et al.* Topological states in engineered atomic lattices. *Nature Physics*. 2017; **13**(7): 668-671.

[21] Li X, Li Q, Ji T *et al.* Lieb Lattices Formed by Real Atoms on Ag (111) and Their Lattice Constant-Dependent Electronic Properties. *Chinese Physics Letters*. 2022; **39**(5): 057301.

[22] Mukherjee S, Spracklen A, Choudhury D *et al.* Observation of a localized flat-band state in a photonic Lieb lattice. *Physical review letters*. 2015; **114**(24): 245504.

[23] Vicencio RA, Cantillano C, Morales-Inostroza L *et al.* Observation of localized states in Lieb photonic lattices. *Physical review letters*. 2015; **114**(24): 245503.

[24] Taie S, Ichinose T, Ozawa H *et al.* Spatial adiabatic passage of massive quantum particles in an optical Lieb lattice. *Nature Communications*. 2020; **11**(1): 257.

[25] Jiang W, Huang H, Liu F. A Lieb-like lattice in a covalent-organic framework and its Stoner ferromagnetism. *Nature communications*. 2019; **10**(1): 2207.

[26] Cui B, Zheng X, Wang J *et al.* Realization of Lieb lattice in covalent-organic frameworks with tunable topology and magnetism. *Nature communications*. 2020; **11**(1): 66.

[27] Jiang W, Zhang S, Wang Z *et al.* Topological band engineering of Lieb lattice in Phthalocyanine-based metal–organic frameworks. *Nano letters*. 2020; **20**(3): 1959-1966.

[28] Feng H, Liu C, Zhou S *et al.* Experimental realization of two-dimensional buckled Lieb lattice. *Nano Letters*. 2020; **20**(4): 2537-2543.

[29] Mann JB, Meek TL, Allen LC. Configuration Energies of the Main Group Elements. *Journal of the American Chemical Society*. 2000; **122**(12): 2780-2783.

[30] Luo Y, Han Y, Liu J *et al.* A unique van Hove singularity in kagome superconductor CsV$_{3-x}$Ta$_x$Sb$_5$ with enhanced superconductivity. *Nature Communications*. 2023; **14**(1): 3819.

[31] Liu L, Wang C, Zhang L *et al.* Surface Van Hove Singularity Enabled Efficient Catalysis in Low-Dimensional Systems: CO Oxidation and Hydrogen Evolution Reactions. *The Journal of Physical Chemistry Letters*. 2022; **13**(3): 740-746.

[32] Zheng Y, Slade TJ, Hu L *et al.* Defect engineering in thermoelectric materials: what have we learned? *Chemical Society Reviews*. 2021; **50**(16): 9022-9054.

[33] Tersoff J, Hamann DR. Theory of the scanning tunneling microscope. *Physical Review B*. 1985; **31**(2): 805.

[34] Xiao J, Yan B. First-principles calculations for topological quantum materials. *Nature reviews physics*. 2021; **3**(4): 283-297.

[35] Soumyanarayanan A, Reyren N, Fert A *et al.* Emergent phenomena induced by spin–orbit coupling at surfaces and interfaces. *Nature*. 2016; **539**(7630): 509-517.

[36] Gao W, Zhu X, Zheng F *et al.* A possible candidate for triply degenerate point fermions in trigonal layered PtBi$_2$. *Nature communications*. 2018; **9**(1): 3249.

[37] Schröter NB, Pei D, Vergniory MG *et al.* Chiral topological semimetal with multifold band crossings and long Fermi arcs. *Nature Physics*. 2019; **15**(8): 759-765.

[38] Liu Y, Li Y, Rajput S *et al.* Tuning Dirac states by strain in the topological insulator $Bi_2Se_3$. *Nature Physics*. 2014; **10**(4): 294-299.
[39] Hu Y, Wu X, Yang Y *et al.* Tunable topological Dirac surface states and van Hove singularities in kagome metal $GdV_6Sn_6$. *Science Advances*. 2022; **8**(38): eadd2024.
[40] Zhang Y, He K, Chang C-Z *et al.* Crossover of the three-dimensional topological insulator $Bi_2Se_3$ to the two-dimensional limit. *Nature Physics*. 2010; **6**(8): 584-588.
[41] Leng P, Chen F, Cao X *et al.* Gate-tunable surface states in topological insulator β-$Ag_2Te$ with high mobility. *Nano Letters*. 2020; **20**(10): 7004-7010.
[42] Wu Q, Zhang S, Song H-F *et al.* WannierTools: An open-source software package for novel topological materials. *Computer Physics Communications*. 2018; **224**: 405-416.
[43] Soluyanov AA, Vanderbilt D. Computing topological invariants without inversion symmetry. *Physical Review B*. 2011; **83**(23): 235401.
[44] Kresse G, Furthmüller J. Efficient iterative schemes for ab initio total-energy calculations using a plane-wave basis set. *Physical review B*. 1996; **54**(16): 11169.
[45] Blöchl PE. Projector augmented-wave method. *Physical review B*. 1994; **50**(24): 17953.
[46] Perdew JP, Burke K, Ernzerhof M. Generalized gradient approximation made simple. *Physical review letters*. 1996; **77**(18): 3865.
[47] Mostofi AA, Yates JR, Lee Y-S *et al.* wannier90: A tool for obtaining maximally-localised Wannier functions. *Computer physics communications*. 2008; **178**(9): 685-699.

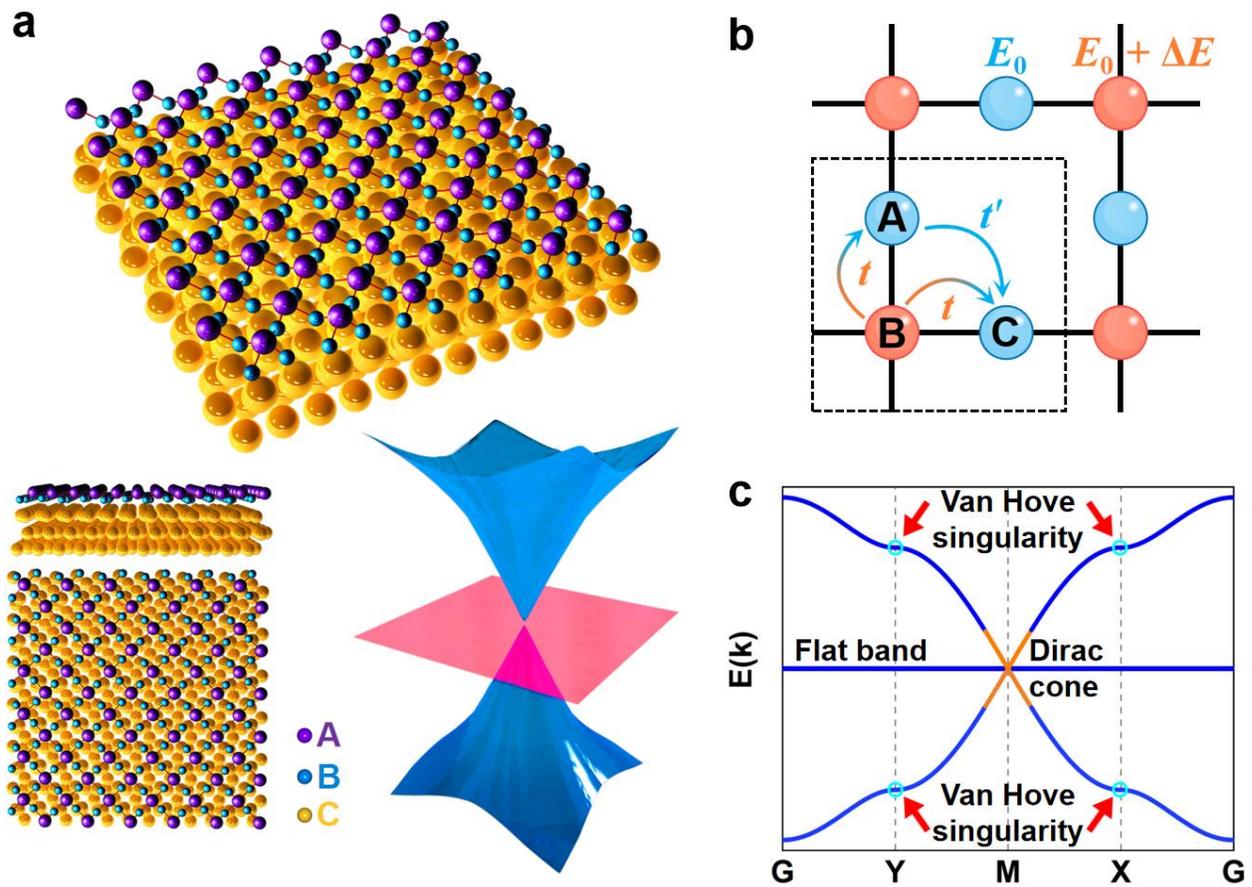

**Figure 1.** Basic design and tight-binding model of the Lieb lattice. (a) Schematic diagram of a Lieb lattice design. (b) Crystal structure of the Lieb lattice with one corner (B) and two edge-centered sites (A and C), where on-site energies are $E_0 + \Delta E$ and $E_0$ for B and A/C, respectively. $t$ and $t'$ represent the nearest-neighbor and next-nearest-neighbor hopping, respectively. The black dashed box indicates the unit cell. (c) Band structure of the Lieb lattice with nearest-neighbor electron hopping. The FB, Dirac cone, and VHSs are clearly seen.

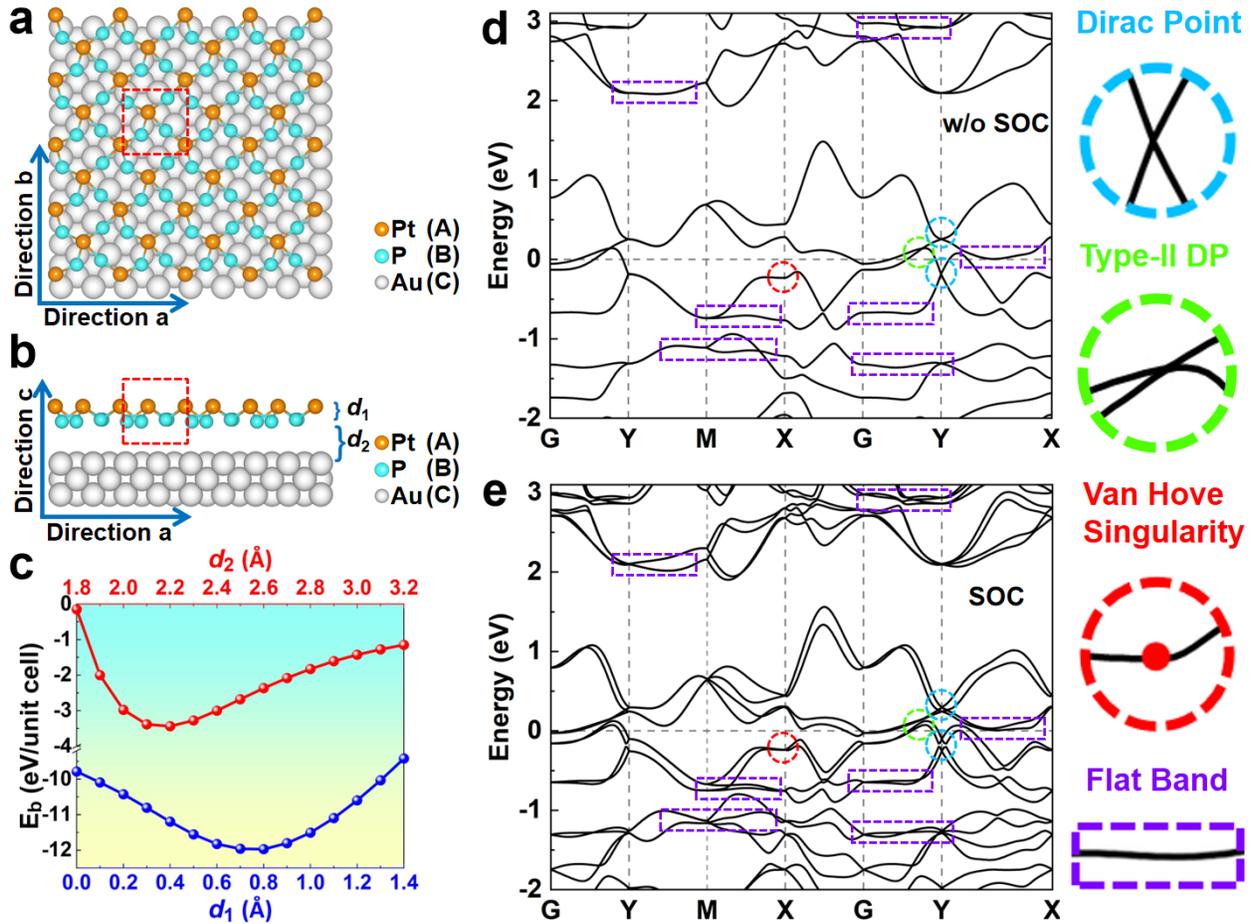

**Figure 2.** Crystal and electronic band structure. (a and b) Top view (a) and side view (b) of the 2D Lieb lattice composed of a √2 × √2 Pt and a 1 × 1 P square-like lattice on a Au(111) surface. The red dashed box indicates the unit cell. Orange/cyan/gray spheres represent Pt/P/Au atoms, respectively. (c) Interlayer binding energy of the 2D Lieb lattice as a function of the interlayer distances indicated in panel (b). (d and e) Calculated electronic band structure of the 2D Lieb lattice without (d) and with (e) spin-orbit coupling. Dirac points, type-II Dirac points, VHSs, and FBs are marked by blue, green, red, and purple dashed lines, respectively.

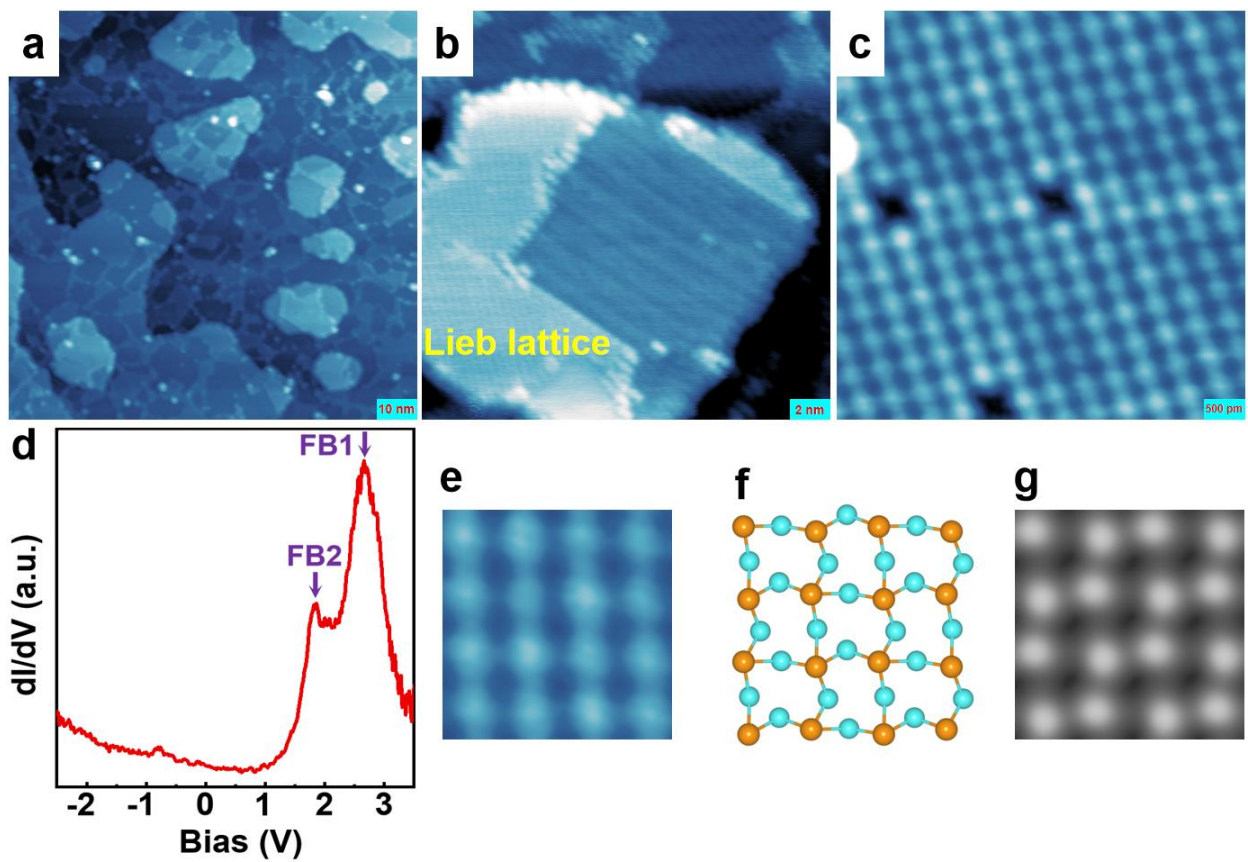

**Figure 3.** Lieb lattices imaged by STM and electronic FB peaks measured by STS. (a) Sample after deposition of Pt atoms and annealing at 300 °C for 80 min. (b) A close-up STM image, Lieb lattices exist in the bright region. (c) Atomically resolved STM image of Lieb lattices. (a) $V_s$ = -1 V, 100 × 100 nm$^2$; (b) $V_s$ = -10 mV, 20 × 20 nm$^2$; (c) $V_s$ = -300 mV, 5 × 5 nm$^2$. (d) The d$I$/d$V$ spectrum for the sites within the Lieb lattice as a function of bias voltage $V$, showing two FB peaks at ~ +2.66 V (FB1) and ~ +1.83 V (FB2), respectively. (e to g) Zoomed-in STM image (e), corresponding atomic model (f), and simulated STM image (g) of Lieb lattices.

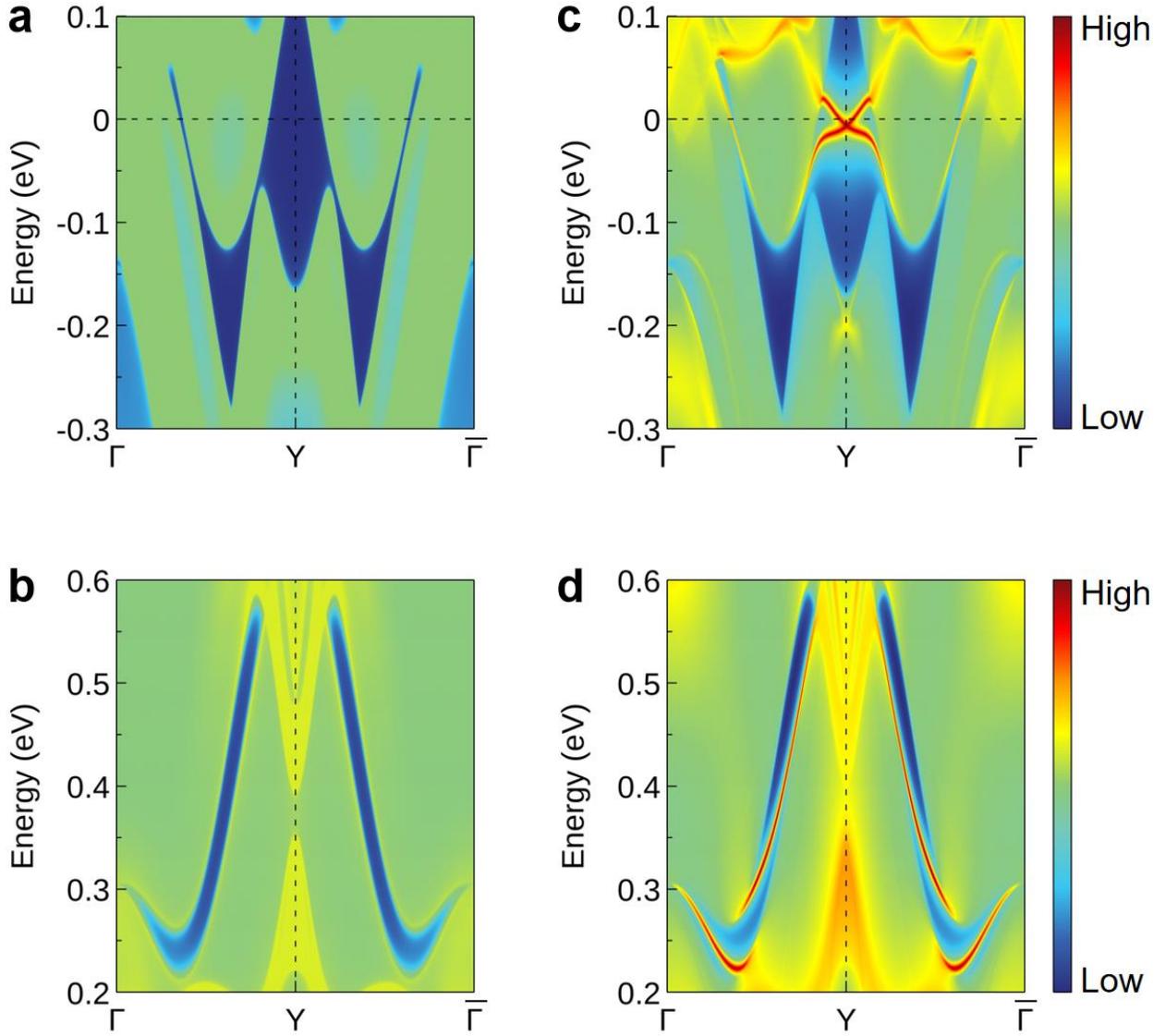

**Figure 4.** Prediction of topological edge states in the Pt-P Lieb lattice. (a and b) Calculated spectral function of the (100) surface states along $\Gamma - Y - \bar{\Gamma}$ paths below (a) and above (b) $E_F$. (c and d) Calculated spectral function of the (100) surface states and edge states along $\Gamma - Y - \bar{\Gamma}$ paths below (c) and above (d) $E_F$. The red contours denote the edge states.

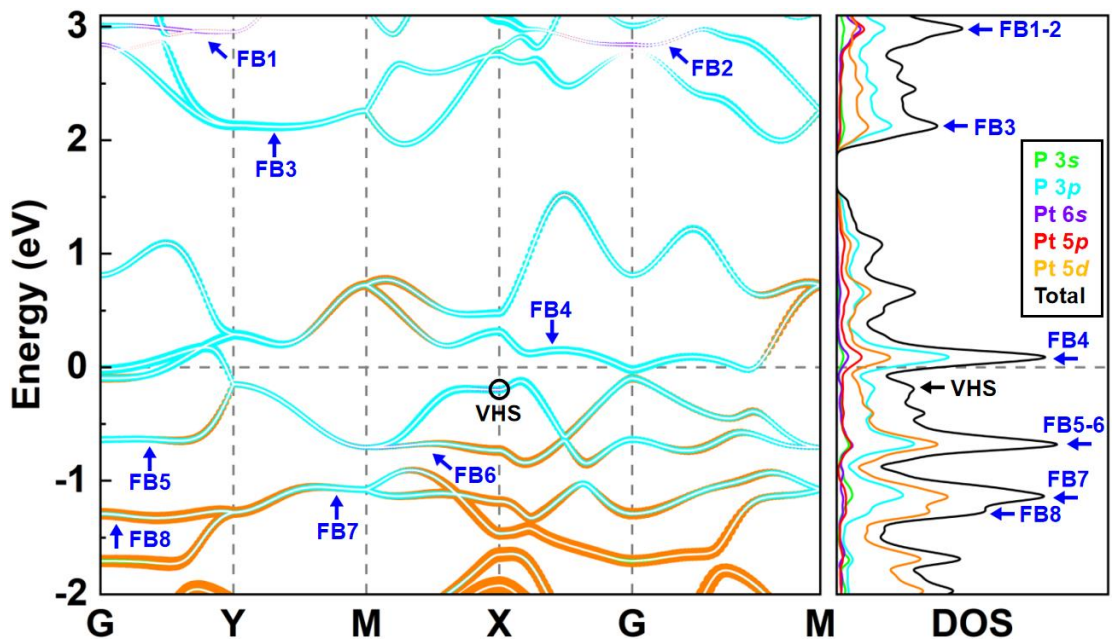

**Figure S1.** Electronic band structure of the Pt-P Lieb lattice. Projected band structure and projected density of state (PDOS) in the Pt-P Lieb lattice. Eight flat bands (FB1-8) and corresponding PDOS peaks are marked by blue arrows. The van Hove singularity is marked by a black circle and the corresponding PDOS peak is marked by a black arrow.

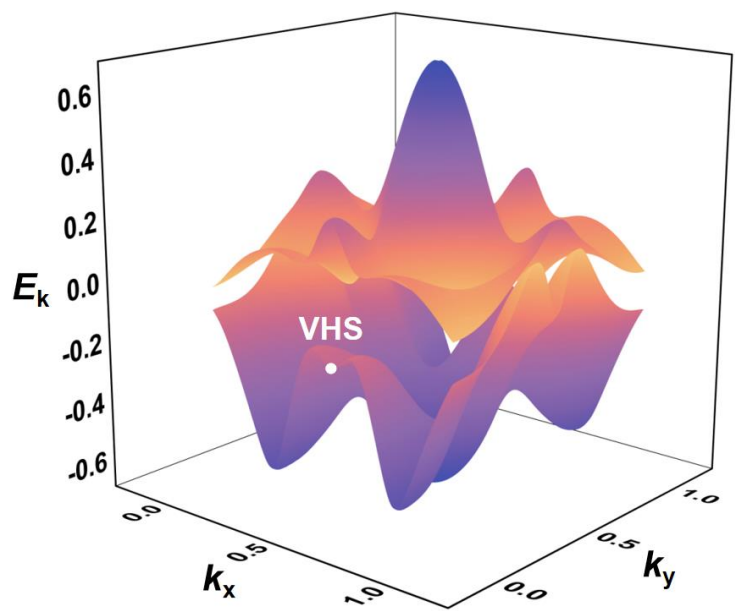

**Figure S2.** Schematic of a van Hove singularity (VHS) in the Pt-P Lieb lattice. Three-dimensional band structure near the Fermi level. The VHS is marked with a white dot.

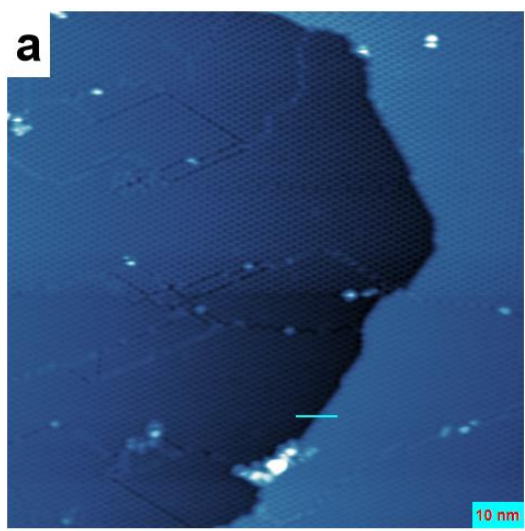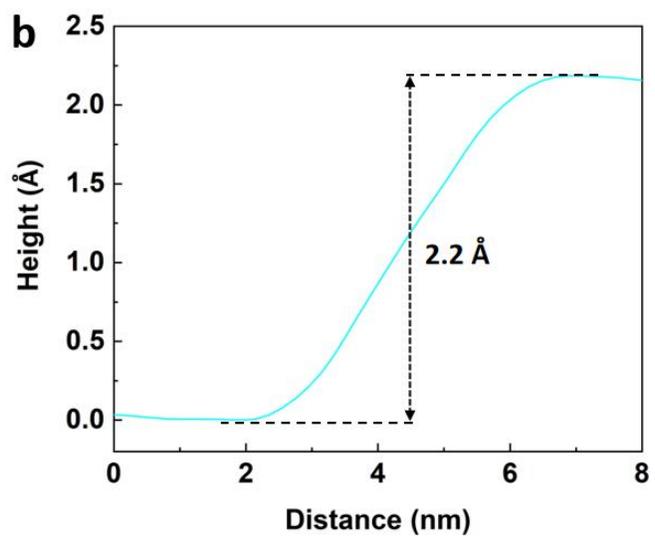

**Figure S3.** P atoms deposited on Au(111). (a) The scanning tunneling microscopy (STM) image of P atoms on a Au(111) surface. (b) Height profile along the cyan line in (a).

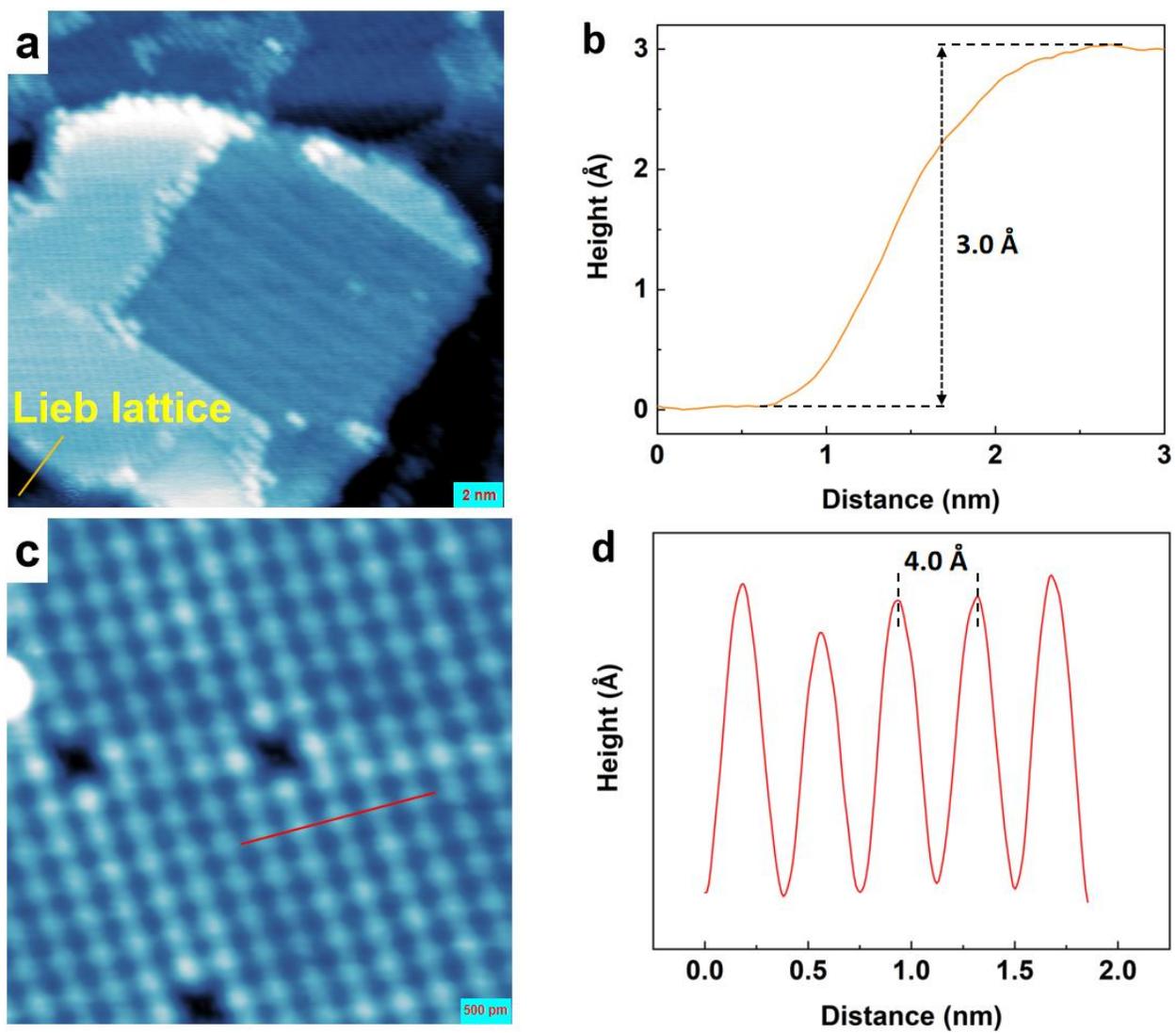

**Figure S4.** Pt-P Lieb lattices deposited on Au(111). (a) The close-up STM image of the Pt-P Lieb lattice on a Au(111) surface. (b) Height profile along the orange line in (a). (c) Atomically resolved STM image of Lieb lattices. (d) Line profile along the red line in (c).

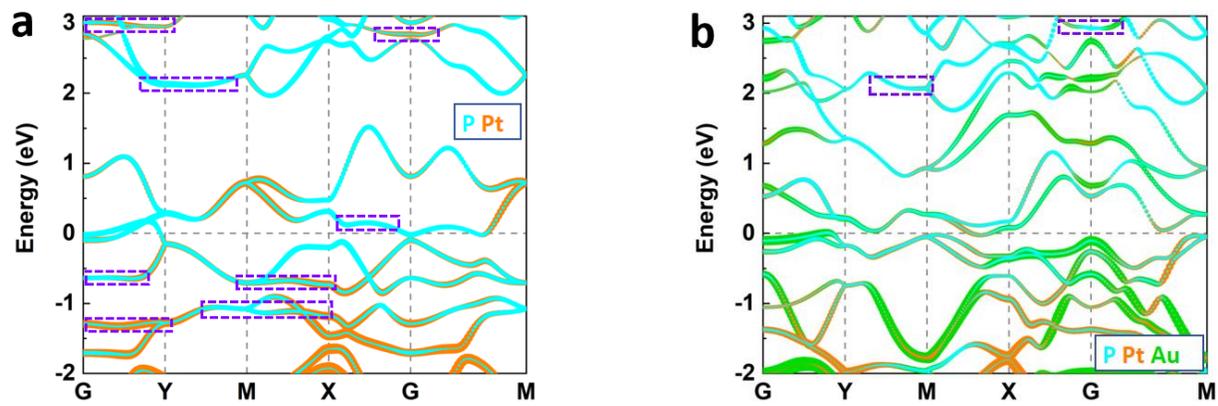

**Figure S5.** Electronic band structure of the Pt-P Lieb lattice. (a and b) Projected band structure of the Pt-P Lieb lattice without a Au substrate (a) and with a single-layer Au substrate (b). Flat bands are highlighted within dashed purple boxes. Contributions from P, Pt, and Au are represented by cyan, orange, and green bands, respectively.

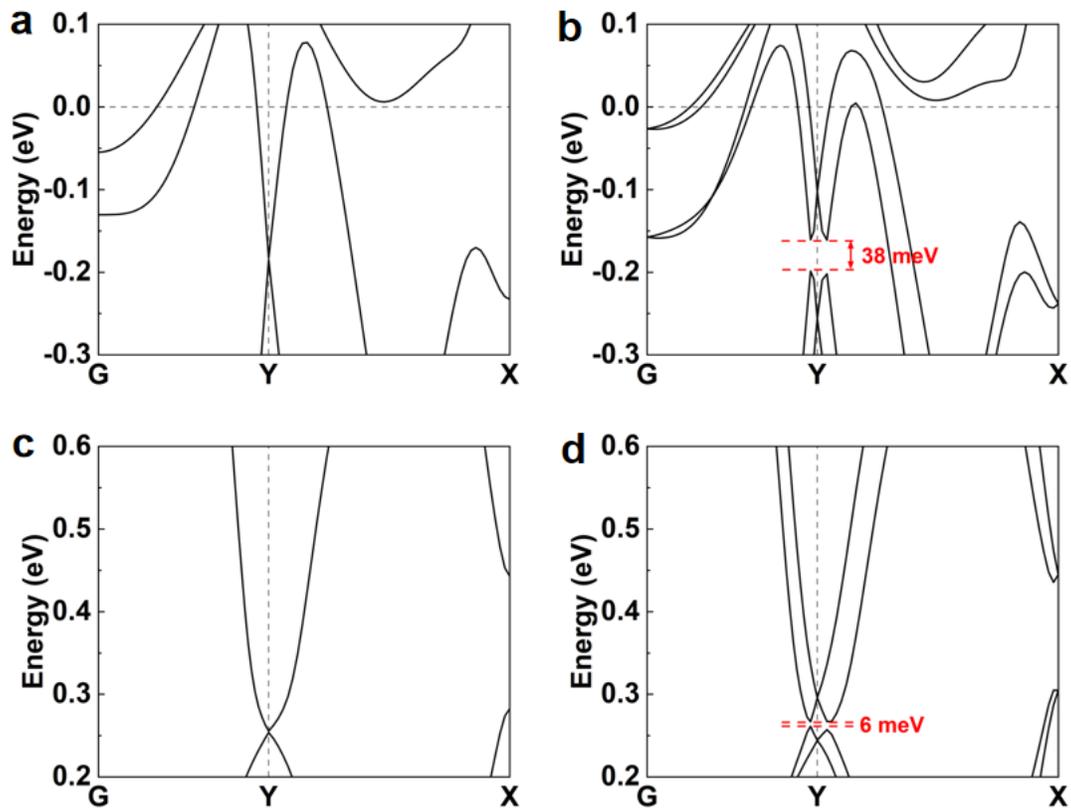

**Figure S6.** Magnified band structure of the Pt-P Lieb lattice. (a and b) Calculated band structure of the Pt-P Lieb lattice without (a) and with (b) spin-orbital coupling, where a bandgap opens at the Dirac cone located at ~$E_F$ - 0.18 eV by ~38 meV with spin-orbital coupling. (c and d) Calculated band structure of the Pt-P Lieb lattice without (c) and with (d) spin-orbital coupling, where a bandgap opens at the Dirac cone located at ~$E_F$ +0.26 eV by ~6 meV with spin-orbital coupling.

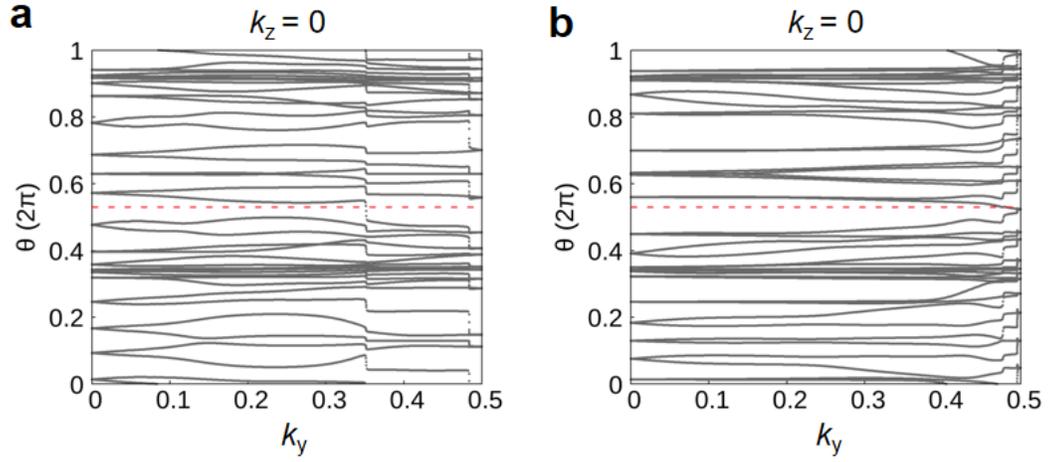

**Figure S7.** $Z_2$ topological invariants for the Pt-P Lieb lattice. (a and b) Evolution of Wannier charge centers (WCCs) for (a) $k_z = 0$ (corresponding to Fig. 4c), (b) $k_z = 0$ (corresponding to Fig. 4d) time-reversal invariant momentum planes. For the two-dimensional Lieb system, it can only take the $Z_2$ number at $k_z = 0$ plane. WCC evolution lines cross an arbitrary reference line (red dashed line) with an odd number of times on the $k_z = 0$ plane, resulting in $Z_2 = 1$.